\documentclass[aps,prd,showpacs,showkeys,superscriptaddress,onecolumn]{revtex4}
\usepackage{epsfig}
\usepackage{graphicx}
\usepackage{amsmath,amssymb,amsfonts,latexsym}
\usepackage{graphicx}

\usepackage{epsfig,amssymb,amsfonts,verbatim}

\usepackage{amsmath}
\usepackage{latexsym}
\usepackage{amsfonts}
\usepackage{amssymb}
\usepackage{color}

\def\bfl{\begin{flushleft}}
\def\efl{\end{flushleft}}
\def\bfr{\begin{flushright}}
\def\efr{\end{flushright}}
\def\bc{\begin{center}}
\def\ec{\end{center}}

\def\ba{\begin{eqnarray}}
\def\ea{\end{eqnarray}}
\def\baa#1{\begin{array}{#1}}
\def\eaa{\end{array}}
\def\bw{\begin{widetext}}
\def\ew{\end{widetext}}
\def\nn{\nonumber }

\def\text#1{\mbox{#1}}

\begin{document}


\title{Magneto-resistivity model and ionization energy approximation for ferromagnets}

\author{Andrew Das Arulsamy}
\email{andrew@physics.usyd.edu.au}
\affiliation{School of Physics, The
University of Sydney, Sydney, New South Wales 2006, Australia}
\affiliation{Institute of Mathematical Sciences, University of Malaya, 50603 Kuala-Lumpur, Malaysia}

\author{Xiangyuan Cui}
\affiliation{School of Physics, The University of Sydney, Sydney, New South Wales 2006, Australia}

\author{Catherine Stampfl}
\affiliation{School of Physics, The University of Sydney, Sydney, New South Wales 2006, Australia}

\author{Kurunathan Ratnavelu}
\affiliation{Institute of Mathematical Sciences, University of Malaya, 50603 Kuala-Lumpur, Malaysia}

\date{\today}

\begin{abstract}
The evolution of resistivity versus temperature ($\rho(T)$) curve for different doping elements, and in the presence of various defects and clustering are explained for both diluted magnetic semiconductors (DMS) and manganites. Here, we provide unambiguous evidence that the concept of ionization energy ($E_I$), which is explicitly associated with the atomic energy levels, can be related quantitatively to transport measurements. The proposed ionization energy model is used to understand how the valence states of ions affect the evolution of $\rho(T)$ curves for different doping elements. We also explain how the $\rho(T)$ curves evolve in the presence of, and in the absence of defects and clustering. The model also complements the results obtained from first-principles calculations. 
\end{abstract}


\maketitle

\section{Introduction}

Ferromagnets have the tremendous potential for the development of spintronics and subsequently will lay the foundation to realize quantum computing. The field of spintronics require the incorporation of the spin-property of the electrons into the existing charge
transport devices~\cite{igor}. Parallel to this, the technological potential of DMS (Ref.~\cite{munekata1}) is associated to spintronics-device development, whereas manganites that show a large drop of resistance below $T_C$ lead to the colossal magnetoresistance effect (CMR), which is also important in the new technologies such as read/write heads for high-capacity magnetic storage and spintronics~\cite{gub}. As such, applications involving both DMS and manganites very much depend on our understanding of their transport properties at various
doping levels and temperatures ($T$). In addition, DMS also has several interesting physical properties namely, anomalous Hall-effect~\cite{mus}, large magnetoresistance in low dimensional geometries~\cite{kana}, the changes of electron-phase-coherence time in the presence of magnetic impurities~\cite{sami} and negative bend resistance~\cite{jaya}. As for the transport properties, there are several models developed
to characterize the resistivity of DMS. In particular, the impurity band model coupled with the multiple exchange
interactions for $T < T_C$ for Ga$_{1-x}$Mn$_x$As was proposed~\cite{esch13}. The electronic states of the impurity band can be either localized
or delocalized, depending on doping concentration or the Fermi-level ($E_F$). If $E_F$ is in the
localized-state, then the conduction is due to carrier hopping. If $E_F$ is in the extended-state, then the conduction is metallic and finite even for $T = 0$ (Ref.~\cite{esch13}). On the other hand, the spin disorder scattering resistivity as a function of magnetic susceptibility can be used to estimate the magnitude of $J_{ex}$ (the ferromagnetic (FM) exchange
interaction energy)~\cite{t-omiya}. Moreover, there are also theories that qualitatively explain the
conductivity for $T > 0$, namely,
the Kohn-Luttinger kinetic exchange model~\cite{lutt} and the semiclassical
Boltzmann model~\cite{jung2,hwang,lopez}. 

Apart from that, for manganites, the one- and two-orbital models~\cite{sen} and the
phase separated resistivity model~\cite{mayr,dietl} have been used to qualitatively 
describe the resistivity curves for $T > 0$. However, in all these approaches, we are faced with two crucial problems, the need (i) to explain how the resistivity evolve with different doping elements, without any \textit{a priori} assumption on carrier density and (ii) to understand how defects and clustering affect the evolution of $\rho(T)$ curves. Here, we show unequivocally, a new method to analyse the evolution of $\rho(T)$ curves for different doping elements using the concept of the $E_I$ invoked in the Hamiltonian and Fermi-Dirac statistics. In doing so, we can also understand the evolution of $\rho(T)$ curves in the presence of defects and clustering, which is important for characterization of spintronics devices. The
$E_I$ concept has broad applications, where it has
been applied successfully for the normal state (above critical temperature) of high temperature
superconductors~\cite{arulsamy2,arulsamy3,arulsamy7} and
ferroelectrics~\cite{arulsamy8}. The $E_I$ model is for compounds obtained via substitutional doping,
not necessarily homogeneous or defect-free. 

\section{Ionization energy model}

\subsection{Carrier density}

A typical solid contains 10$^{23}$ strongly interacting particles. Therefore,
their universal collective behavior is of paramount interest as
compared to the microscopic details of each particular particle and the
potential that surrounds it. This universal collective behavior,
being the focal point in this work, arises out of Anderson's
arguments in \textit{More is Different}.~\cite{anderson} That is,
we intend to justify a universal physical parameter that
could be used to describe the association between the
transport-measurement data and the fundamental properties of an
atom. In view of this, we report here the existence of such a parameter through the Hamiltonian as given below (Eq.~(\ref{eq:100})). The parameter is the ionization energy, a macroscopic, many-electron atomic parameter.  

\begin {eqnarray}
\hat{H}\varphi = (E_0 \pm \xi)\varphi, \label{eq:100}
\end {eqnarray}

where $\hat{H}$ is the usual Hamilton operator and $E_0$ is the total energy at $T$ = 0. The + sign of $\pm\xi$
is for the electron ($0 \rightarrow +\infty$) while the $-$ sign
is for the hole ($-\infty \rightarrow 0$). Here, we define the ionization energy in a crystal, $\xi = E_I^{\rm{real}}$ is approximately proportional to $E_I$ of an isolated atom or ion. Now, to prove the validity of Eq.~(\ref{eq:100}) is quite easy because $\xi$ is also an eigenvalue and we did not touch the Hamilton operator. Hence, we are not required to solve Eq.~(\ref{eq:100}) in order to prove its validity. We can prove by means of constructive (existence) and/or direct proofs, by choosing a particular form of wavefunction with known solution (harmonic oscillator, Dirac-delta and Coulomb potentials) and then calculate the total energy by comparison. In doing so, we will find that the total energy is always given by $E_0 \pm \xi$, as it should be (see Appendix and Ref.~\cite{andrew}). For an isolated atom, the concept of ionization energy implies that (from Eq.~(\ref{eq:100}))

\begin {eqnarray}
&&\pm \xi = E_{\rm{kin}} - E_0 + V_{\rm{pot}} = \pm E_I, \nn
\end {eqnarray}

where $E_I$ is the ionization energy of an isolated atom. The corresponding total energy is $E_0 \pm \xi = E_{\rm{kin}} + V_{\rm{pot}} = E_0 \pm E_I$. Whereas for an atom in a crystal, the same concept of ionization
energy implies that $\pm \xi = E_{\rm{kin}} - E_0 + V_{\rm{pot}} + V_{\rm{many-body}} = E_I
+ V_{\rm{many-body}} = \pm E^{\rm{real}}_I$. Here, $V_{\rm{many-body}}$ is the many body potential averaged from the
periodic potential of the lattice. The corresponding total energy is $E_0 \pm \xi = E_{\rm{kin}} + V_{\rm{pot}} + V_{\rm{many-body}} = E_0
\pm E_I + V_{\rm{many-body}} = E_0 \pm E^{\rm{real}}_I$. Here, $E_I^{\rm{real}}$ is the ionization
energy of an atom in a crystal. The exact values of $E_I$ are known
for an isolated atom. That is, one can still use $E_I$ obtained from isolated atoms for $r \rightarrow \infty$ in order to predict the evolution of resistivity versus temperature curves for different doping elements. Therefore,
Eq.~(\ref{eq:100}) can be approximately rewritten as $\hat{H}\varphi \propto (E_0 \pm E_I)\varphi$. It is obvious from the Hamiltonian given in Eq.~(\ref{eq:100}) that we cannot use it to determine interactions responsible for FM
and the magnitude of $T_C$, simply because we have suppressed the $V_{\rm{many-body}}$. On the other hand, if we have a free-electron system, then the total energy equation is given by $E = E_0 \pm \sum_i^z\sum_j E_{Ii,j}^{\rm{real}} = E_0 \pm \beta \sum_j E_{Ij}$, where, $E_0$ is the total energy of the compound at $T$ = 0 and
$j$ is the sum over the constituent elements in a particular compound. We also define here, $\beta = 1 + \frac{\langle V(x)\rangle}{E_I}$, where $\langle V(x)\rangle$ is the averaged many-body potential value. Here, $E_I^{\rm{real}}$ can be noted as one of the many-body response functions with respect to transport properties. The total energy can be rewritten as 

\begin {eqnarray}
&E& = E_0 \pm \sum_i^z\sum_j E_{Ii,j}^{\rm{real}} \nn \\&& = E_0 \pm
[E_{\rm{kin}} - E_0 + V_{\rm{pot}} + V_{\rm{many-body}}] \nn \\&& = E_{\rm{kin}} +
V_{\rm{pot}} + V_{\rm{many-body}} \Leftrightarrow ~\rm{for~
electrons}~ \pm \rightarrow + \nn \\&& = E_{\rm{kin}} + V_{\rm{total}} \nn \\&& = E_{\rm{kin}} \Leftrightarrow \rm{implies~free~electrons}. \nn 
\end {eqnarray}

\begin{figure}[hbtp!]
\begin{center}
\scalebox{0.3}{\includegraphics{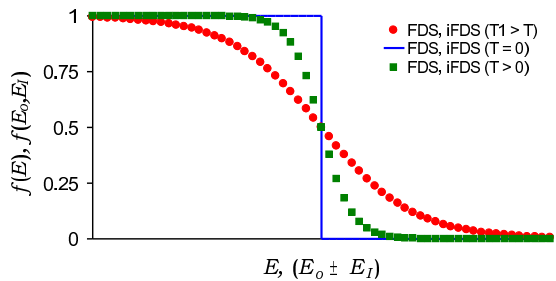}}
\caption{Standard Fermi Dirac (FDS) and ionization energy based Fermi-Dirac ($i$FDS) distributions for temperatures, $T$ = 0, $T >$ 0 and $T1 > T$.}
\label{fig:1}
\end{center}
\end{figure}

Note that we have substituted for $E_I^{\rm{real}}$ since the
concept of ionization energy is irrelevant here simply because the
electrons in these metals are free and do not require excitations
from its parent atom to conduct electricity. As such, the carrier
density is constant and independent of temperature. Whereas, the
scattering rate is the one that determines the resistivity with
respect to temperature, impurities, defects, electron-electron and
electron-phonon interactions. Therefore, the total energy from Eq.~(\ref{eq:100}) carries the
\textit{fingerprint} of each constituent atom in a compound and it refers to the difference in the
energy levels of each atom rather than the absolute values of each
energy level in each atom. Hence, the kinetic energy of
each electron from each atom will be captured by the total energy and preserves the atomic level
\textit{electronic-fingerprint} in the compound. Parallel to Eq.~(\ref{eq:100}), the electron and hole distribution functions can be derived as~\cite{arulsamy2,arulsamy3} (see Fig.~\ref{fig:1})

\begin{eqnarray}
&&f_e(E_0,E_I) = \frac{1}{e^{[\left(E_0 + E_I
\right) - E_F^{(0)}]/k_BT }+1}, \nn
\\&& f_h(E_0,E_I) = \frac{1}{e^{[E_F^{(0)} - \left(E_0 - E_I
\right)]/k_BT}+1}. \label{eq:360}
\end{eqnarray}

where $E_F^{(0)}$ is the Fermi level at $T$ = 0, independent of doping concentration and $k_B$ is the Boltzmann constant. From Eq.~(\ref{eq:360}), one can surmise that large $E_I$ corresponds
to the difficulty of the electron to conduct due to a large Coulomb
attraction between the electron and the ionic core. This effect
will give rise to low conductivity. The variation of $E_I$ with doping in our model
somewhat resembles the variation produced from the impurity band model with doping~\cite{esch13,burch}. The
difference is that our model fixes the Fermi level as a constant and we let $E_I$ capture all the changes due to
doping, consistent with ionization energy based Fermi-Dirac statistics ($i$FDS). Whereas, in the impurity-band
approach, both the impurity band and the
Fermi level evolve simultaneously with doping, consistent with
Fermi-Dirac statistics. 

The carrier density can be calculated from

\begin{eqnarray}
n = \int_0^\infty{f_e(E_0,E_I)N_e(E_0)dE_0}, \label{eq:3600}
\end{eqnarray} 

where $N_e(E_0)$ is the density of states and $E_F$ is defined from now
on as the Fermi level at $T$ = 0 so as to comply with
Eq.~(\ref{eq:100}). 

\subsection{Spin-orbit coupling}

Here, we will show the relationship of the ionization energy to the energy level splitting and the spin-orbit coupling, as well as their association with resistivity. The latter association is crucial because spin-orbit coupling is found to be an an important phenomenon that influences the electronic properties of the magnetic semiconductors~\cite{cou,kaes}. It is well established that the energy associated with both the relativistic correction and the spin-orbit coupling for Hydrogen-like atoms is given by the fine structure formula, $E_{fs}$ (Ref.~\cite{beth})

\begin{eqnarray}
E_{fs} = -\frac{E_\texttt{n}}{\texttt{n}^2}\bigg[1 + \frac{(\alpha_{fs} Z)^2}{\texttt{n}}\bigg(\frac{1}{k} - \frac{3}{4\texttt{n}}\bigg)\bigg], \label{eq:999}
\end{eqnarray}

where $\alpha_{fs}$ is the fine structure constant, $k$ = $j$ + 1/2, $j$ = $l$ $\pm$ 1/2, where $j$ is the total angular momentum, $l$ is the orbital angular momentum and $E_\texttt{n}$ is the non-relativistic energy level. $Z$ and $\texttt{n}$ are the atomic and the principal quantum numbers, respectively. Using Eqs.~(\ref{eq:100}) and~(\ref{eq:999}), we can show the extreme magnitude of the energy level splitting (i.e., between $k$ = 1 and $k$ = $\texttt{n}$) is 

\begin{eqnarray}
\Delta E_{fs} = (\alpha_{fs} Z)^2(E_0 \pm \xi)_\texttt{n}\frac{(\texttt{n}-1)}{\texttt{n}^2}. \label{eq:9999}
\end{eqnarray}

\begin{figure}[hbtp!]
\begin{center}
\scalebox{0.45}{\includegraphics{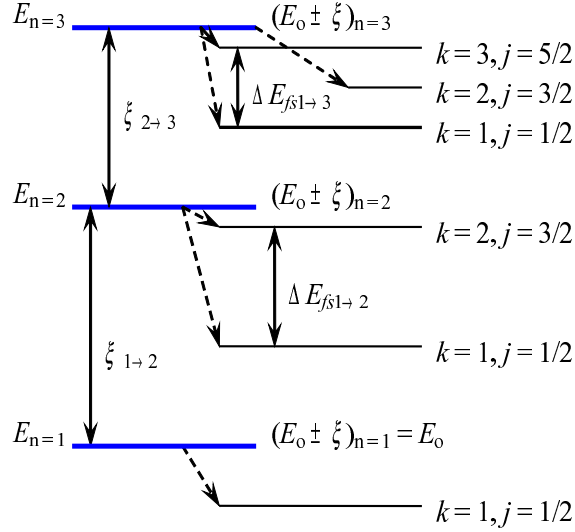}}
\caption{Energy levels of hydrogen-like atom and the energy level splitting (not to scale). $E_{\texttt{n}=1,2,3}$ is the standard energy level notation while, $(E_0 \pm \xi)_{\texttt{n}=1,2,3}$ is the new notation introduced from Eq.~(\ref{eq:9999}).}   
\label{fig:2}
\end{center}
\end{figure}

Thus, the magnitude of the energy level splitting ($\Delta E_{fs}$) is proportional to the ionization energy ($\xi$) and $Z^2$, while it is inversely proportional to $\texttt{n}$. That is, $\Delta E_{fs}$ also decreases with decreasing ionization energy (see Fig.~\ref{fig:2}). Therefore, we can surmise that a system with large ionization energy gives rise to large spin-orbit coupling, which is important for spin-injection~\cite{murakami} in $p$-type DMS. As discussed earlier however, a large ionization energy also leads to a small carrier density (from Eq.~(\ref{eq:3600})), which in turn implies that spin-orbit coupling competes with the electrical conductivity. In addition, Fig.~\ref{fig:2} points out that $\Delta E_{fs 1 \rightarrow 2}$ and $\xi_{1 \rightarrow 2}$ are larger than $\Delta E_{fs 1 \rightarrow 3}$ and $\xi_{2 \rightarrow 3}$, respectively, in accordance with Eqs.~(\ref{eq:999}) and~(\ref{eq:9999}). Therefore, we can surmise that $\Delta E_{fs 1 \rightarrow 2}$ $\propto$ $\xi_{1 \rightarrow 2}$, $\Delta E_{fs 1 \rightarrow 3}$ $\propto$ $\xi_{2 \rightarrow 3}$, $\Delta E_{fs 1 \rightarrow 4}$ $\propto$ $\xi_{3 \rightarrow 4}$ and so on.

\subsection{Resistivity model based on ionization energy}

In order to derive the resistivity model as a function of ionization energy, we need an expression that connects the carrier density with the total current. As a consequence, we propose that the total current in ferromagnets consists of contributions from both
non-FM and FM phases, which is $J$ = $\sum_\nu J_\nu$, with $\nu$ =
$e^{\downarrow}$, $se^{\uparrow}$, $sde^{\downarrow}$, $h^{\downarrow}$,
$sh^{\uparrow}$, $sdh^{\downarrow}$. Here, $se^{\uparrow}$ and $sh^{\uparrow}$ are the spin-assisted electron and hole, respectively below $T_C$, influenced by the $T$-dependent magnetization function ($M(T,M_0)$) through the spin disorder scattering rate ($\tau_{SD}$). $e^{\downarrow}$ and $h^{\downarrow}$ are the electron and hole with electron-electron scattering rate ($\tau_{e-e}$). $sde^{\downarrow}$ and $sdh^{\downarrow}$ are the electron and hole with spin disorder scattering and $M(T,M_0)$ = constant (Const.) for $T
> T_C$. For convenience, the spin-up, $\uparrow$ denotes
the direction of the magnetic field or a particular direction
below $T_C$, while the spin-down, $\downarrow$ represents any
other direction. That is, a non-FM phase gives rise to $J_{e,h}$ and $J_{sde,sdh}$. Whereas, $J_{se,sh}$ originates from a FM phase where ($J_{sde,sdh}$,$J_{e,h}$) $\rightarrow$ $J_{se,sh}$ for $T
< T_C$. Hence, this resistivity model is fundamentally different from the phase separation model~\cite{mayr}. That is, $J_{e,h}
(\propto \tau_{e-e}$) $<$ $J_{se,sh} (\propto \tau_{SD}[M(T,M_0)]$) for $T <
T_C$, while $J_{e,h} (\propto \tau_{e-e}$) and $J_{sde,sdh} (\propto \tau_{SD}$[Const.]) contribute for $T
> T_C$ due to the \textit{non-existence} of the FM phase. In other words, for $T > T_C$, the whole
system is a non-FM phase and its conductivity is determined by
$J_{e,h} (\propto \tau_{e-e}$) and $J_{sde,sdh} (\propto \tau_{SD}$[Const.]). For $T < T_C$, some of the non-FM
phase becomes FM due to spin and its conductivity is determined by $J_{e,h}
\propto \tau_{e-e}$ and $J_{se,sh} (\propto \tau_{SD}[M(T,M_0)]$). Simply put, we assume that $\tau_{e-e}$ and $\tau_{SD}$[Const.] contribute in the non-FM phase above $T_C$. Below $T_C$, we have both FM and non-FM phases contributing to the resistivity via $\tau_{e-e}$ and $\tau_{SD}[M(T,M_0)]$. As such, the total current can be simplified
as $J$ = $J^{\downarrow}_e$ + [$J^{\uparrow}_{se}$,$J^{\downarrow}_{sde}$] = $J_e$ +
[$J_{se}$,$J_{sde}$] if the considered system is $n$-type, while $J$ = $J_h$
+ [$J_{sh}$,$J_{sdh}$] if it is $p$-type. $J_e$ and $J_h$ are the spin
independent charge current due to electron and hole, respectively, and are influenced by $\tau_{e-e}$. $J_{sde}$ and $J_{sdh}$ are also the spin
independent charge current due to electron and hole, respectively, but are influenced by $\tau_{SD}$[Const.] in the
non-FM phase, which is valid only above $T_C$. $J_{se}$ and $J_{sh}$ are the spin-assisted
charge current in the FM phase only, influenced by the $\tau_{SD}[M(T,M_0)]$. Thus, the total resistivity (for $n$ or
$p$-type) can be written as (after making use of the elementary
resistivity equation, $\rho = m/ne^2\tau$)

\begin{eqnarray}
&\rho^{-1}& = \rho^{-1}_{e,h} + \big[\rho^{-1}_{sde,sdh},\rho^{-1}_{se,sh}\big] = \bigg[\frac {m_{e,h}^*}{(n,p)e^2 \tau_{e-e}}\bigg]^{-1} \nn \\&& + \bigg[\frac {m_{e,h}^*}{(n,p)e^2 \tau_{SD}\big[{\rm{Const.}},M(T,M_0)\big]}\bigg]^{-1} \nn \\&& = \bigg[\frac {m_{e,h}^*}{(n,p)e^2 \tau_{e-e}}\bigg]^{-1} + \bigg[\frac {m_{e,h}^*}{(n,p)e^2 \tau_{SD}}\bigg]^{-1}, \label{eq:4}  
\end{eqnarray}

where $m_{e,h}^*$ denotes the effective mass of the electron or hole.
$e$ is the charge of an electron. The carrier density for the
electron and hole ($n,p$) based on $i$FDS  are given
by (after substituting Eq.~(\ref{eq:360}) into Eq.~(\ref{eq:3600}))

\begin{eqnarray}
n = 2\left[\frac{k_BT}{2\pi\hbar^2}\right]^{3/2}(m^*_e)^{3/2}
\exp\left[\frac{E_F - E_I}{k_BT}\right], \label{eq:5}
\end{eqnarray}

\begin{eqnarray}
p = 2\left[\frac{k_BT}{2\pi\hbar^2}\right]^{3/2}(m^*_h)^{3/2}
\exp\left[\frac{-E_F - E_I}{k_BT}\right]. \label{eq:6}
\end{eqnarray}

The spin disorder scattering resistivity as derived by
Tinbergen-Dekker is given by~\cite{tinbergen3}

\begin {eqnarray}
&&\rho_{SD}(T<T_C) = \frac{(m^*_{e,h})^{5/2}N(2E_F)^{1/2}}{\pi
(n,p) e^2\hbar^4}J_{ex}^2 \nn \\&& \times \bigg[S(S+1) -
S^2\bigg(\frac{M_{TD}(T)}{M_0}\bigg)^2 -
S\bigg(\frac{M_{TD}(T)}{M_0}\bigg) \nn \\&& \times \tanh
\bigg(\frac{3T_CM_{TD}(T)}{2TS(S+1)M_0}\bigg)\bigg], \label{eq:3}
\end {eqnarray}

where $M_{TD}(T,M_0)$ is the Tinbergen-Dekker magnetization
function~\cite{tinbergen3}. $M_0$ is the magnetization at zero temperature. $N$ is the concentration of nearest neighbor ions (for example, the concentration of Mn). $\hbar$ = $h/2\pi$, $h$ denotes Plancks
constant and $S$ is the spin quantum
number. Equation~(\ref{eq:3}) is equal to the theory developed by
Kasuya~\cite{kasuya4} if one replaces the term,
$\tanh\big[3T_CM_{TD}(T)/2TS(S+1)M_0\big]$ with 1. Substituting
$1/\tau_{e-e}$ = $A_{e,h}T^2$ (due to the electron-electron
interaction and $A_{e,h}$ is the $T$ independent electron-electron scattering rate
constant), together with Eqs.~(\ref{eq:3}) and~(\ref{eq:5})(or~(\ref{eq:6}))
into Eq.~(\ref{eq:4}), then one can arrive at

\begin{eqnarray}
\rho_{e,se}(T) = \frac{AB\exp \big[(E_I\mp
E_F)/k_BT\big]}{AT^{3/2}[M_{\alpha}(T,M_0)]^{-1}+ BT^{-1/2}},
\label{eq:7}
\end{eqnarray}

\begin{eqnarray}
&&A =
\frac{A_{e,h}}{2e^2(m^*_{e,h})^{1/2}}\left[\frac{2\pi\hbar^2}{k_B}\right]^{3/2},\nn
\\&& B = \frac{2m^*_{e,h}N(\pi E_F)^{1/2}J_{ex}^2}{e^2\hbar k_B^{3/2}},
\nn \\&& \tau_{SD}^{-1} =
\left[\frac{N(2E_F)^{1/2}(m^*_{e,h})^{3/2}}{\pi
\hbar^4}\right]J_{ex}^2M_{\alpha}(T,M_0). \nn
\end{eqnarray}

Note here that we have invoked the strong correlation through the carrier density, which is a function of the real ionization energy. Secondly, to account for different spin polarization below and above $T_C$, we have $n$ = $n^{\uparrow} + n^{\downarrow}$, $1/m^* = 1/m^{*\uparrow} + 1/m^{*\downarrow}$ and $\tau = \tau_{e-e}^{\downarrow} + \tau_{SD}^{\uparrow}(J_{ex})$. Notice that the spin-disorder scattering rate, $\tau_{SD}^{\uparrow}$ is a function of the exchange energy, $J_{ex}$. Therefore, by redefining the variables in the elementary resistivity equation in accordance with strongly correlated effects, we have now derived a resistivity equation (Eq.~(\ref{eq:7})) suitable for strongly correlated ferromagnets. Here, $M_{\alpha}(T,M_0)$ is the $T$-dependent magnetization function for $T < T_C$. For $T > T_C$ however, $M_{\alpha}(T,M_0)$ = constant. The negative and positive signs in $E_I\mp E_F$ are for
electrons and holes, respectively. Equation~(\ref{eq:7}) points out
that, $\rho_{e,se}(T)$ is semiconducting and
it only becomes metallic if $E_I\mp E_F < T$ and the FM phase sets in.
Below $T_C$, the FM metallic phase increases with lowering $T$ and
once it achieves the maximum value (saturation) at a much lower
$T$, insulating character may set in as a result of $E_I\mp E_F >
T$. We define this temperature $T_{\rm{crossover}}$ which corresponds to the transition from FM metallic to insulating below
$T_C$. Hence, it is clear that our model is fundamentally
different to the phase separation model as pointed out earlier.
The emergence of insulating ($T_{\rm{crossover}}$) character below
$T_C$ is an intrinsic property based on the ionization energy
model, regardless of defect densities and it was first predicted
for both high-$T_c$ superconductors and ferromagnets~\cite{arulsamy2}. In a recent analysis using the one-
and two-orbital models, a similar insulating character below $T_C$
was observed from numerical calculations~\cite{sen}, in support of
our prediction. However, the analysis stops short of explaining why
the insulating behavior persists even in the ``clean limit" with no 
defects (i.e. pure host material with zero interstitial and vacancy defects). Here we propose that $T_{\rm{crossover}}$ or the
insulating behavior below $T_C$ is intrinsic and is associated
with the energy levels through the $E_I \mp E_F$ parameter. If $E_I \mp E_F > T
> T_C$, then $\rho_{e,se}(T) \propto \exp[(E_I \mp E_F)/T]$. In this
limit, a large CMR effect could be
observed if the metallic-ferromagnetism sets in. The empirical function of the normalized magnetization is defined here as

\begin{eqnarray}
M_{\rho}(T,M_0) = 1-\frac{M_{\rho}(T)}{M_0}\label{eq:8}.
\end{eqnarray}

Equation~(\ref{eq:8}) is an empirical function that will be used
to extract the magnetization curve from the resistivity curve via 
Eq.~(\ref{eq:7}). In other words, Eq.~(\ref{eq:8}) is used to
calculate the magnetization curve, after coupling it with
Eq.~(\ref{eq:7}). For example, the magnetization curves associated
with Tinbergen-Dekker ($TD$)~\cite{tinbergen3}, Kasuya ($K$)~\cite{kasuya4} and resistivity curves
($\rho$) are calculated using

\begin{eqnarray}
&&M_{TD}(T,M_0) \nn \\&&  = S(S+1) - S^2\bigg(\frac{M_{TD}(T)}{M_0}\bigg)^2 -
S\bigg(\frac{M_{TD}(T)}{M_0}\bigg)\nn \\&& \times
\tanh\bigg[\frac{3T_CM_{TD}(T)}{2TS(S+1)M_0}\bigg],\label{eq:B8}
\end{eqnarray}

\begin{eqnarray}
&&M_{K}(T,M_0) \nn \\&&  = S(S+1) - S^2\bigg(\frac{M_{K}(T)}{M_0}\bigg)^2 -
S\bigg(\frac{M_{K}(T)}{M_0}\bigg),\label{eq:B9}
\end{eqnarray}

and Eq.~(\ref{eq:8}), respectively. That is,
Eqs.~(\ref{eq:B8}),~(\ref{eq:B9}) and~(\ref{eq:8}) are separately
coupled with Eq.~(\ref{eq:7}) to fit the resistivity curves and
subsequently obtain $\frac{M_{TD}(T)}{M_0}$,
$\frac{M_{K}(T)}{M_0}$ and $\frac{M_{\rho}(T)}{M_0}$,
respectively. Consequently, we can compare and analyze $M_{\alpha}(T)/M_0$ with the experimentally measured,
$M_{\rm{exp}}(T)/M_0$ data. Recall that, $\alpha$ = $K$ (calculated from
Eq.~(\ref{eq:B9})), $TD$ (from Eq.~(\ref{eq:B8})), $\rho$ (from
Eq.~(\ref{eq:8})), exp (Experimentally determined
magnetization curves). For simplicity, we will drop the term $M_0$
from the equations that contain $\frac{M_{\alpha}(T)}{M_0}$.

For the electron- or hole-doped strongly correlated non-FM semiconductors, one
needs Eq.~(\ref{eq:9}) given below, which is again based on
$i$FDS,~\cite{arulsamy2,arulsamy3,arulsamy7,arulsamy8}

\begin{eqnarray}
\rho_{e,h} = AT^{1/2} \exp\left[\frac{E_I \mp E_F}{k_BT}\right],
\label{eq:9}
\end{eqnarray}

where the $-$ sign is for the electron and + sign is for the hole. Equation~(\ref{eq:9}) will be used to justify the importance of the term $J_{se}$ even if the resistivity is semiconductor-like in
the FM phase. 

The values of $E_I$ can be averaged using the approximation given below

\begin {eqnarray}
E_I^{\rm{real}} = \beta\sum_i^z \frac{E_{Ii}}{z} \propto E_I = \sum_i^z \frac{E_{Ii}}{z},\label{eq:25}
\end {eqnarray}

where $\beta$ is a function of the many-body potential, $V_{\rm{many-body}}$ and varies with
different ``background" atoms (host lattice). The $i$ $=$ 1, 2,..., $z$ represent the first, second,...etc
ionization energies and, $z$ is the oxidation number of a
particular ion. $E^{\rm{real}}_I$ is actually equal to the energy
needed to ionize an atom or ion in a crystal such that the
electron is excited to a distance $r$. On the other hand, $E_I$ corresponds to taking that particular electron to $r \rightarrow
\infty$ with $V_{\rm{many-body}} = 0$. 

\begin{figure}[hbtp!]
\begin{center}
\scalebox{0.4}{\includegraphics{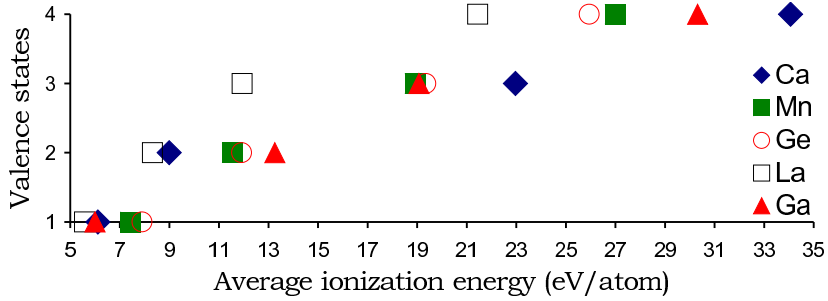}}
\caption{Average ionization energy distribution for different isolated atoms and valence states. Calculated from Ref.~\cite{web28} using Eq.~(\ref{eq:25})}   
\label{fig:3}
\end{center}
\end{figure}

Prior to averaging, the ionization energies for all the elements mentioned above were taken from Ref.~\cite{web28}, also given in Fig.~\ref{fig:3}. We can also use Eq.~(\ref{eq:Z12}) that originates from Eq.~(\ref{eq:25}) to predict the change in the valence state of a particular ion.  

\begin{eqnarray}
&&\frac{\delta}{j}\sum^{z+j}_{i=z+1}{E_{Ii}} +
\frac{1}{z}\sum^{z}_{i=1}{E_{Ii}} =
\frac{1}{q}\sum^{q}_{i=1}{E_{Ii}}. \label{eq:Z12} 
\end{eqnarray}

Here, $\frac{\delta}{j}\sum^{z+j}_{i=z+1}{E_{Ii}}$ has $i$ $=$ $z$
+ 1,..., $z$ + $j$ and $j$ $=$ 1, 2,.... It is solely
due to the multivalence ion. That is, the first term is due to
Mn$^{4+}$ ion's contribution (Mn$^{3+}$ $-$ electron $\to$ Mn$^{4+}$ = 51.200 eV atom$^{-1}$), hence
$j$ equals 1 ($=4-3$) in this case and $\delta$ represents
the additional contribution from Mn$^{4+}$. The second ($i$ $=$ 1,
2,..., $z$) and last ($i$ $=$ 1, 2,..., $q$) terms
respectively are due to Mn $-$ 3(electrons)
$\to$ Mn$^{3+}$ and Ga $-$ 3(electrons) $\to$ Ga$^{3+}$. Recall
that $q$ = $z$ = 3+ and $i$ = 1, 2,... represent the first,
second,... ionization energies while $j$ = 1, 2,...
represent the fourth, fifth,... ionization energies.

\section{Ionization energy model applied to ferromagnets}

\subsection{Ga$_{1-x}$Mn$_x$As}

\subsubsection{$\rho(T > T_C)$}

We will first apply the $E_I$-based carrier density to the resistivity ($\rho(T)$) curves of DMS above $T_C$ by means of $\rho(T > T_C) \propto 1/n$, in order to avoid the influence of fitting parameters in our analysis. Subsequently, we will discuss the evolution of resistivity curves with appropriate fittings. The Ga$_{1-x}$Mn$_x$As system is the most studied material among the DMS with $T_C$ around 170 K (Ref.~\cite{jung2}). The origin of ferromagnetism and the transport properties in this class of materials are still unclear quantitatively~\cite{jung2}. The evolution of the resistivity curve ($\rho(T,x)$) for sample \emph{X1}, Ga$_{1-x}$Mn$_x$As (see Fig.~3 of Ref.~\cite{matsu}) is
such that $\rho(T,0.015)$ $>$ $\rho(T,0.022)$ $>$ $\rho(T,0.035)$ $>$
$\rho(T,0.043)$ $>$ $\rho(T,0.053)$. That is, the $\rho(T,x)$
curve shifts downwards with increasing dopant concentration, $x$. This is expected from the $i$FDS model since $E_I$ for Mn$^{3+}$ (18.910 eV atom$^{-1}$) is
less than $E_I$ for Ga$^{3+}$ (19.070 eV atom$^{-1}$), assuming
Mn$^{3+}$ substitutes Ga$^{3+}$. However, the resistivity curve, $\rho(T,0.071)$ is
above $\rho(T,0.035)$, unexpectedly. 

For sample \emph{Y2} (see Fig.~1 of
Ref.~\cite{oiwa}), $\rho(T,0.015)$ $>$ $\rho(T,0.022)$ $>$
$\rho(T,0.035)$, also complies with the $i$FDS model. 
Again, unexpectedly we have, $\rho(T,0.043)$ $<$
$\rho(T,0.053)$ $<$ $\rho(T,0.071)$, where $\rho(T,x)$ shifts
upwards with increasing $x$. That is, $\rho(T,x)$ switched over from
decreasing (expected) to increasing with $x$ at critical
concentrations, $x_{c1}$ = 0.071 and $x_{c2}$ = 0.043 for samples
\emph{X1} and \emph{Y2}, respectively. These switch-overs that seem to violate
$i$FDS can be explained if we can understand what causes $E_I$ to
deviate from its averaged value. The only reason that could give rise to such deviation is the change in the ions valence states. 

Using Eq.~(\ref{eq:Z12}), we obtain: $[\delta E_I
({\rm{Mn^{4+}}})] + \frac{1}{3}[E_I {\rm{(Mn^{3+})}} + 
E_I {\rm{(Mn^{2+})}} + E_I {\rm{(Mn^{1+})}}] = \frac{1}{3}[E_I ({\rm{Ga^{3+}}}) + E_I ({\rm{Ga^{2+}}}) + E_I
({\rm{Ga^{1+}}})]$ and $(51.200\times \delta) + 18.910 = 19.070$, therefore $\delta = 0.003$. Since
the average $E_I$ (18.910 eV atom$^{-1}$) for Mn$^{3+}$ is less than
the average $E_I$ (19.070 eV atom$^{-1}$) for Ga$^{3+}$, and if the
resistivity curve shifts upward with Mn substitution, then $z$ +
$\delta$ gives the minimum valence number for Mn (Mn$^{>(z+\delta)+}$) which can be calculated from
Eq.~(\ref{eq:Z12}). If however, the resistivity curve shifts
downward with Mn substitution, then $z$ + $\delta$ gives the
maximum valence number for Mn (Mn$^{<(z+\delta)+}$). Consequently, the maximum valence
state for Mn in Ga$_{1-x}$Mn$_x$As is 3.003+ (Mn$^{<3.003+}$) for the case where the $\rho(T,x)$ curve shifted downwards with $x$. If the valence state of Mn is larger than 3.003+, then $\rho(T,x)$ is expected to shift upwards with $x$. Therefore, the switch-over from decreasing to increasing $\rho(T,x)$ with doping, $x$ is due to larger average Mn valence state, and it is calculated to be larger than 3.003+. Next, we need to understand what can cause this change in the average Mn valence state. The most likely reason comes from Mn occupation at non-substitutional sites, i.e., the Mn$^{3+}$ ions do not substitute Ga$^{3+}$ ions. This has been predicted experimentally where the formation of Mn interstitials (Mn$_I$) is found to be substantial above a critical concentration, $x_c$ (Ref.~\cite{esch13}). Interestingly, occupations at interstitial sites is found to change the charge states of Mn ions, as predicted from the first-principles calculations, where Mn$^{5+}$ and Mn$^{4+}$ are more stable at interstitial sites, while Mn$^{3+,2+,1+}$ are stable at substitutional sites~\cite{priya}. In addition, interstitial in the presence of clustering gives rise to larger positive charge states as compared to clustering due to substitutional alone in which, larger positive charge implies larger valence state for Mn ions~\cite{priya}. Therefore, we can understand that at certain doping, the average Mn valence states increased due to Mn$_I$ and clustering and eventually reduces the carrier density and shifts the $\rho(T,x)$ curve upwards with doping.   

\subsubsection{$\rho(T > 0)$}

\begin{figure}[hbtp!]
\begin{center}
\scalebox{0.6}{\includegraphics{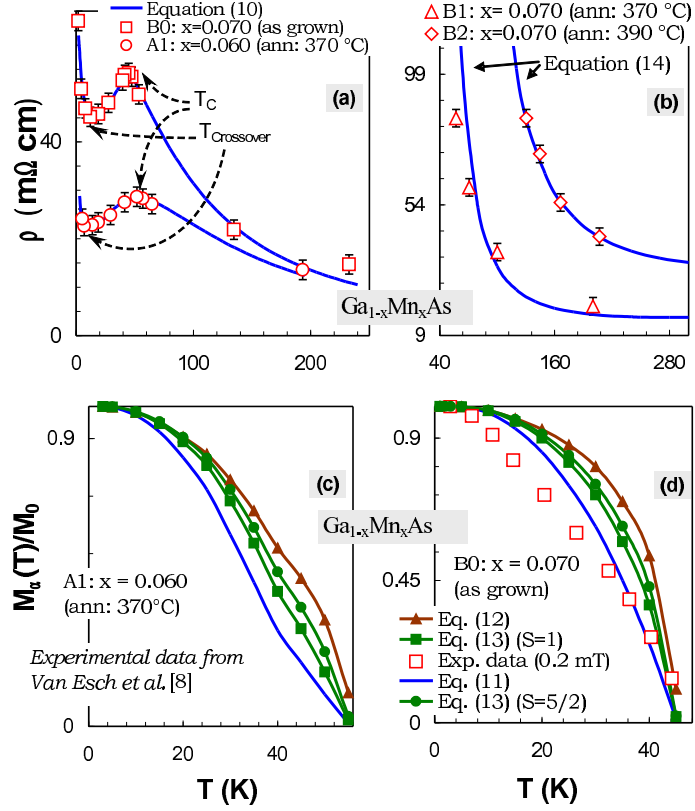}}
\caption{(a) Calculated, using Eq.~(\ref{eq:7}), and experimental resistivity curves ($\rho(T)$) as a function of temperature for Ga$_{1-x}$Mn$_x$As. B0 and A1 indicate Ga$_{0.930}$Mn$_{0.070}$As (as grown) and Ga$_{0.940}$Mn$_{0.060}$As (annealed at 370$^{\circ}$C), respectively. (b) Calculated, using Eq.~(\ref{eq:9}), and experimental $\rho(T)$ curves for annealed
non-ferromagnetic Ga$_{0.930}$Mn$_{0.070}$As samples. B1 and B2 indicate Ga$_{0.930}$Mn$_{0.070}$As (annealed at 370$^{\circ}$C) and Ga$_{0.930}$Mn$_{0.070}$As (annealed at 390$^{\circ}$C), respectively. (c) and (d)
show the $T$ variation of calculated normalized magnetization $M_{\alpha}(T)$ where $\alpha$ = $K$ (Eq.~(\ref{eq:B9})), $TD$ (Eq.~(\ref{eq:B8})), $\rho$ (Eq.~(\ref{eq:8})), exp (experimentally determined
magnetization data) curves with spin quantum number $S$ = 1 for $x$ = 0.060 and 0.070
respectively. $M_K(T)$ is also calculated with $S$ = 5/2 as shown in (c) and (d). The
experimental results $M_{\rm{exp}}(T)$ for $x$ = 0.070 (as grown) are
shown in (d).}
\label{fig:4}
\end{center}
\end{figure}

Now we will apply the ionization energy incorporated resistivity model to the resistivity measurements~\cite{esch13} and the fits based
on Eqs.~(\ref{eq:7}) and~(\ref{eq:9}) are shown in
Figs.~\ref{fig:4}(a) and \ref{fig:4}(b) respectively for Ga$_{1-x}$Mn$_x$As.
One needs two fitting parameters ($A$ and $E_I \mp E_F$) for
$\rho(T>T_C)$ and another two ($B$ and $M_{\rho}(T,M_0)$) for
$\rho(T<T_C)$. All the fitting parameters are listed in
Table~\ref{Table:I}. 


\begin{table}[ht!]
\caption{Calculated values of $A$, $B$ and the ionization energy
($E_I$). ``Ann. $T$" denotes the annealing temperature and $T_{cr}$ = $T_{crossover}$, is the temperature that corresponds to the transition from FM metallic to insulating character. ``Calc." denotes calculated values from the fits.} 
\begin{tabular}{l c c c c c} 
\hline\hline 
\multicolumn{1}{l}{Samples} & Ann. $T$ (\textbf{H})   &     $A$              & $B$ & $E_I \mp E_F$ (Calc.)   & $T_C$ ($T_{cr}$)    \\  
\multicolumn{1}{l}{}        & $^oC$ (Tesla)         &    (Calc.)        & (Calc.)  &   Kelvin (meV)  &   Kelvin            \\
\hline 
Ga$_{0.940}$Mn$_{0.060}$As$^{(a)}$ & 370 (0)      & 4.5  & 400  & 8 (0.69)   & 50 (10) \\ 
Ga$_{0.930}$Mn$_{0.070}$As$^{(a)}$ & As grown (0) & 9.2  & 400  & 12 (1.04)  & 45 (12) \\
Ga$_{0.930}$Mn$_{0.070}$As$^{(a)}$ & 370 (0)      & 0.02 & -    & 280 (24.2) & -       \\
Ga$_{0.930}$Mn$_{0.070}$As$^{(a)}$ & 390 (0)      & 0.03 & -    & 400 (34.5) & -       \\
La$_{0.9}$Ca$_{0.1}$MnO$_3$$^{(b)}$ & - (0)       & 10   & 0.65 & 1400 (121) & 222 (-) \\ 
La$_{0.8}$Ca$_{0.2}$MnO$_3$$^{(b)}$ & - (0)       & 10   & 1.2  & 1300 (112) & 246 (-) \\ 
La$_{0.8}$Ca$_{0.2}$MnO$_3$$^{(b)}$ & - (6)       & 5    & 3.2  & 900 (78)   & 251 (-) \\ [1ex] 
\hline 
\end{tabular}
\footnotetext{(a) Ref.~\cite{esch13}} 
\footnotetext{(b) Ref.~\cite{mahendiran2}} 
\label{Table:I} 
\end{table}

Note that $S$ = 1 and 5/2 are used for the
fits of $M_K(T)$ while $T_C$ and $T_{crossover}$ ($T_{cr}$)
are determined from the experimental resistivity curves. The
$T_{cr}$'s observed in Ga$_{0.940}$Mn$_{0.060}$As (annealed:
370$^o$C) and Ga$_{0.930}$Mn$_{0.070}$As (as grown) are 10 K and
12 K, respectively, which are very close to the calculated values
of 8 K and 12 K, respectively from Eq.~(\ref{eq:7}). The
calculated carrier density is 2 $\times$ 10$^{20}$ cm$^{-3}$,
using $E_I \mp E_F$ $\sim$ 10 K, and Eq.~(\ref{eq:5}). In this
calculation, we take the effective mass as $m^*_e$ = 10$m_0$, in
accordance with the optical spectroscopy measurements~\cite{burch}
and compensation due to annealing, where $m_0$ is the rest mass of
an electron.

Figures~\ref{fig:4}(c) and \ref{fig:4}(d) show the normalized
magnetization, $M_{\alpha}(T)$. $M_{\rho,TD,K}(T)$ are compared
with the experimentally determined magnetization~\cite{esch13}
($M_{\rm{exp}}(T)$) as depicted in Fig.~\ref{fig:4}(d). One can
easily notice the inequality, $M_{TD}(T)$ $>$ $M_{K}(T)$ $>$
$M_{\rho}(T)$ $>$ $M_{\rm{exp}}(T)$ from Figs.~\ref{fig:4}(c) and \ref{fig:4}(d).
As such, $M_{\rho}(T)$ is the best fit as compared with
$M_{\rm{exp}}(T)$, better than the models developed by
Tinbergen-Dekker~\cite{tinbergen3} and Kasuya~\cite{kasuya4}.

\subsection{Mn$_x$Ge$_{1-x}$}

\subsubsection{$\rho(T > T_C)$}

Another DMS material that we will consider here is the Mn$_x$Ge$_{1-x}$ system. It is a $p$-type DMS, with carrier density of
the order of $10^{19}-10^{20}$ cm$^{-3}$ for 0.006 $\leq$ $x$
$\leq$ 0.035 at room temperature~\cite{park3}. From the $E_I$ model, Mn$^{3+}$ (18.910 eV atom$^{-1}$) substitution into Ge$^{4+}$
(25.941 eV atom$^{-1}$) sites will shift the $\rho(T)$ curve downwards
since ($E_I$)$_{\rm{Ge^{4+}}}$ $>$ ($E_I$)$_{\rm{Mn^{3+}}}$. This
has been observed experimentally (see Fig. 2B of Ref.~\cite{park3}) where,
$\rho(T,x=0.009)$ $>$ $\rho(T,x=0.016)$ $>$ $\rho(T,x=0.02)$.
Subsequently, we can estimate the maximum valence state for the Mn ion
in Mn$_x$Ge$_{1-x}$ using Eq.~(\ref{eq:Z12}) and we find,
$\delta = 0.137$ or Mn$^{<3.137+}$: $[\delta E_I ({\rm{Mn}}^{4+})] + \frac{1}{3}[E_I
({\rm{Mn}}^{3+}) + E_I ({\rm{Mn}}^{2+}) + E_I ({\rm{Mn}}^{1+})] = \frac{1}{4}[E_I
({\rm{Ge}}^{4+}) + E_I ({\rm{Ge}}^{3+}) + E_I ({\rm{Ge}}^{2+}) + E_I ({\rm{Ge}}^{1+})]$.
Therefore, ($51.200\times \delta) + 18.910 = 25.941$ and $\delta = 0.137$. If the
valence state of Mn is larger than 3.137+, then the $\rho(T,x)$ curve
is expected to shift upwards with $x$. The result, $\rho(T,x = 0.02) \approx \rho(T,x = 0.033)$ indicates that Mn$^{3+}$ in this doping range may not have substituted Ge, instead it could have occupied non-substitutional sites that eventually causes the change in the valence state of Mn. Since the resistivity, $\rho(T,x = 0.02) \approx \rho(T,x = 0.033)$, we can calculate the maximum increment of Mn$^{4+}$ content from $x$ = 0.02 to $x$ = 0.033. Using $\delta$ = 0.137, we estimate that there are 13.7\% more Mn$^{4+}$ in Mn$_{0.033}$Ge$_{0.966}$ as compared to Mn$_{0.02}$Ge$_{0.98}$ samples.

\subsubsection{$\rho(T > 0)$}

\begin{figure}[bthp!]
\begin{center}
\scalebox{0.6}{\includegraphics{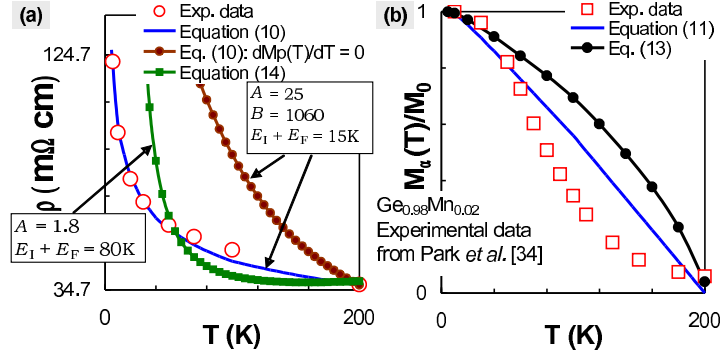}}
\caption{(a) Calculated, using Eq.~(\ref{eq:7}), and experimental resistivity curve $\rho(T)$, for Mn$_{0.02}$Ge$_{0.98}$. Also shown is the calculated $\rho(T)$ curve with the additional constraint, $dM_{\rho}(T)/dT$ = 0 in Eq.~(\ref{eq:7}) to emphasize the influence of $M_{\rho}(T)$ for an
accurate fit. For these calculation, $A$ = 25, $B$ = 1060 and
$E_I+E_F$ = 15 K. We also show that the $T$-dependence of $\rho(T)$ as calculated using Eq.~(\ref{eq:9}) is in accordance with charge current ($J_e$) only. This result lacks the ability to capture the
experimental result. In this case, we used $A$ = 1.8 and
$E_I+E_F$ = 15 K. Both $E_I+E_F$ = 15 K and $E_I+E_F$ = 80 K give
$p$ of the order of 10$^{19}$ cm$^{-3}$ using Eq.~(\ref{eq:5}) and
$m^*_h$ = $m_0$ (rest mass). (b) shows the calculated and experimental $T$ dependence of the magnetization curves $M_{\alpha}(T)$ where $\alpha$ = $K$ (Eq.~(\ref{eq:B9})), $\rho$ (Eq.~(\ref{eq:8})), exp (experimentally determined
magnetization data). Notice the inequality,
$M_{\rho}(T)$ $>$ $M_{\rm{exp}}(T)$ that arises as a result of the
principle of least action.}
\label{fig:5}
\end{center}
\end{figure}

Resistivity measurements~\cite{park3} and the fit using Eq.~(\ref{eq:7})
are shown in Fig.~\ref{fig:5}(a). From the fit, we find that
$E_I \mp E_F$ = 15 K for Mn$_{0.02}$Ge$_{0.98}$ and the hole
density is 2.38 $\times$ 10$^{19}$ cm$^{-3}$ using
Eq.~(\ref{eq:5}), where we take $m_h^*$ = $m_0$, as the lower
limit. This lower limit value is still comparable with the
experimental value given above. We also find that the
semiconductor-like behavior of $\rho(T,x=0.02)$ below $T_C$ is not
exponentially driven as indicated by Eq.~(\ref{eq:7}) in
Fig.~\ref{fig:5}(a).

The pronounced effect of Eq.~(\ref{eq:8}) can be noticed by
comparing the calculated plots between Eq.~(\ref{eq:7}), and
Eq.~(\ref{eq:7}) with the additional constraint $dM_{\rho}(T)/dT$ = 0,
as depicted in Fig.~\ref{fig:5}(a). The normalized
magnetization, $M_{K,\rho,\rm{exp}}(T)$ for Mn$_{0.02}$Ge$_{0.98}$ is
given in Fig.~\ref{fig:5}(b). Again, we find that $M_{\rho}(T)$
gives the best fit for the experimental data, which is better than
$M_{K}(T)$ and $M_{TD}(T)$. However, $M_{\rho}(T)$ is
significantly larger than $M_{\rm{exp}}(T)$, which makes the fit poor. The reason is that the resistivity measures only the lowest
$E_I$ path, regardless of the temperature and with easily-aligned
spin path that gives rise to high conductivity and complies with
the principle of least action. That is, the ability of both $J_e$
and $J_{se}$ to follow the easiest path. On the contrary,
magnetization measurements quantify the average spin of an
ensemble of electrons, which will be smaller in magnitude as
compared with transport measurements.       

\subsection{La$_{1-x}$Ca$_x$MnO$_3$}

\subsubsection{$\rho(T > T_C)$}

For manganites, the La$_{1-x}$Ca$_x$MnO$_3$ system has a maximum $T_C$ of 260 K (Ref.~\cite{mahendiran2}). Unlike DMS, the resistivity of this class of materials above $T_C$ is exponential and thus gives rise to the large drop in resistance below $T_C$ that leads to the CMR effect. We will evaluate this scenario with the ionization energy concept and show the validity of $i$FDS in both DMS and manganites. Based on $E_I$ model, Ca$^{2+}$ ($E_I$ = 9.000 eV atom$^{-1}$) $<$ La$^{3+}$ ($E_I$ = 11.940
eV atom$^{-1}$), therefore the resistivity $\rho(T,x)$ curve is expected to shift
downward with Ca$^{2+}$ doping. On the contrary, the overall $\rho(T)$ curve between $x$ = 0.1 and 0.2, above $T_C$ are almost identical (see Fig.~3 of Ref.~\cite{mahendiran2}). Again, this could be due to the change in the valence state of Mn, as a result of Mn, Ca and/or La occupying the non-substitutional sites, provided that the valence states of Ca$^{2+}$ and La$^{3+}$ are invariant to doping and defects. Indeed,
the content of Mn$^{4+}$ is found to increase with Ca
doping~\cite{mahendiran2}, from 19\% (for $x$ = 0.1) to 25\% (for $x$ =
0.2). We can calculate the maximum increment of Mn$^{4+}$ from $x$ = 0.1 to
0.2, and we obtain, 5.75\% using Eq.~(\ref{eq:Z12}): $[\delta \times E_I ({\rm{Mn}}^{4+})] +
\frac{1}{2}[E_I ({\rm{Ca}}^{2+}) + E_I ({\rm{Ca}}^{1+})] =
\frac{1}{3}[E_I ({\rm{La}}^{3+}) + E_I ({\rm{La}}^{2+}) + E_I ({\rm{La}}^{1+})]$.
Therefore, $(51.200 \times \delta ) + 9.000 = 11.940$ and $\delta = 0.0575$. This value is
remarkably close to the experimental value of 6\%(=25\%$-$19\%),
determined via redox titrations~\cite{mahendiran2}. This implies that, $E_I^{\rm{real}} \propto E_I$, which in turn implies that the averaged many-body or lattice potential ($\beta$) can be approximated as a constant.
This comes as no surprise because the averaged crystal or lattice
potential is indeed a constant due to the periodicity. 

\subsubsection{$\rho(T > 0)$}

\begin{figure}[bthp!]
\begin{center}
\scalebox{0.6}{\includegraphics{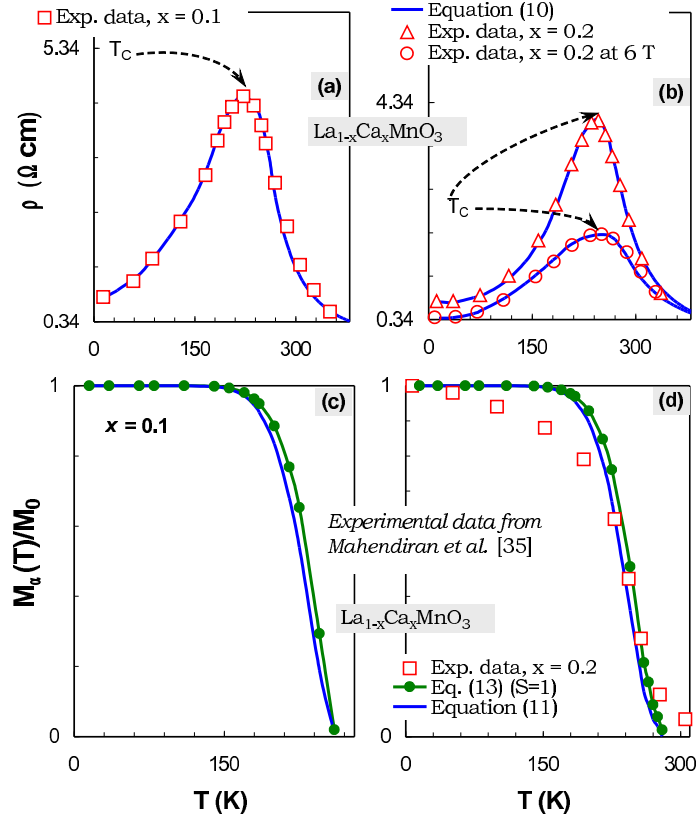}}
\caption{(a) shows the experimental results for the resisitivity versus temperature, $\rho(T)$, for
La$_{1-x}$Ca$_x$MnO$_3$ for $x$ = 0.1, 0.2 and 0.2 (6 Tesla) and (b) the calculated $\rho(T)$ using Eq.~(\ref{eq:7}). All
fits are indicated with solid lines. (c) and (d) show
the $T$ variation of the magnetization, $M_{\alpha}(T)$, where $\alpha$ = $K$ (Eq.~(\ref{eq:B9})) and $\rho$ (Eq.~(\ref{eq:8})) with $S$ = 1 for $x$ = 0.1 and 0.2, respectively. The
experimental $M_{\rm{exp}}(T)$ curve for $x$ = 0.2 is given in (d).}
\label{fig:6}
\end{center}
\end{figure}

Using Eq.~(\ref{eq:7}) for La$_{1-x}$Ca$_x$MnO$_3$~\cite{mahendiran2}, $E_I \mp E_F$ is calculated for $x$ = 0.1
and 0.2 samples and we obtain 0.121 eV (1400 K) and 0.112 eV (1300 K) respectively. The calculated carrier density, using $m^*$ = $m_0$
and Eq.~(\ref{eq:5}) gives 10$^{17}$ cm$^{-3}$. In the presence of
the magnetic field, {\bf H} = 6 Tesla, we obtain $E_I \mp E_F$ =
0.0776 eV for $x$ = 0.2 and its hole concentration $p$ = 10$^{18}$ cm$^{-3}$. The
fits are shown in Figs.~\ref{fig:6}(a) and \ref{fig:6}(b), while the fitting
parameters are listed in Table~\ref{Table:I}. The $\delta$ value determined previously (5.75 \%) needs to be corrected because in the previous calculation, we used $(E_I\mp E_F)_{x=0.1} \approx (E_I\mp E_F)_{x=0.2}$
that gives maximum $\delta$. Therefore, the actual increment after the fitting is
$\delta = 5.75 \times (0.112 \rm{eV}/0.121 \rm{eV}) = 5.32\%$.

Figures~\ref{fig:6}(c) and \ref{fig:6}(d) depict the calculated $M_{\alpha}(T)$
with $S$ = 1 and $M_{\rm{exp}}(T)$ for $x$ = 0.2, respectively. It can be seen that the
magnetization curve calculated from Eq.~(\ref{eq:8}) shows better agreement with the experimental data than $M_K(T)$. Hence, the ionization energy model is
suitable for both types of ferromagnets, be it diluted or
concentrated. However, this does not imply that the ferromagnetic
interactions are identical between diluted and concentrated
ferromagnets. 

\section{Conclusions}

In conclusion, we have developed a theoretical model based on the ionization energy concept that can be used to analyze the evolution of the 
resistivity versus temperature curves of both diluted and concentrated ferromagnets, for different doping elements. By identifying the cause that deviate the ionization energy from its averaged value, we come to understand how defects and clustering contribute to the changes in valence states of ions and eventually how they affect the resistivity of ferromagnets. 

\section*{\textbf{Acknowledgments}}

A.D.A. is grateful to the School of Physics, University of Sydney for the USIRS award, and Kithriammah Soosay for the partial financial support. Special thanks to A. Stroppa
for his explanation on the half-metallic character of MnGe and MnSi. X.Y.C. and C.S. gratefully acknowledge support from
the Australian Research Council (ARC). K.R. acknowledges partial funding from the Malaysian grant No. SAGA 66-02-03-0077. Author-contributions; A.D.A. designed the overall structure of the theory, developed and explained all the ideas related to the theory with proofs, carried out all the analysis and wrote both the manuscript and the Appendix; X.Y.C and C.S. contributed to the idea that the valence states in the ionization energy theory can be related to the First-Principles charge states (in section III-A-1), and edited the manuscript; K.R. edited the manuscript. 

\section{Appendix}

\subsection{Additional Notes}

The relation between the electron affinity and ionization energy is through $r$, where the electron affinity is connected to $r \rightarrow \infty$, and for any other finite $r$, electron affinity is undefined microscopically and it is not a good quantum mechanical variable, as opposed to our $\xi$. For example, if we were to work with the electron affinity then we will face two problems, firstly, considering the anions, electrons from different cations will be treated with equal affinity toward the anion, which is incorrect. Secondly, considering the cations, there is no mathematical analogy as the first electron affinity, second, etc., which we have used for first, second, etc. ionization energies. Hence, the electron affinity in our case, is not a good many-body variable, except for qualitative descriptions. As for the reason behind the name, ionization energy is that in the early stages of the ionization energy theory, we used the atomic ionization energy ($r \rightarrow \infty$) as the input parameter to compute carrier concentrations~\cite{arulsamy2}, and for this reason, $\xi$ was labeled as the ionization energy. The technically correct label is the excitation energy. We can also explain the resistivity above the Curie temperature with any exponential function that either contains the activation energy or the energy related to the electrons hopping rates. However, the whole point of our theory is to associate our exponential function (as given in Eq.~(\ref{eq:7})) with each constituent atoms that exist in a given compound. This enable us to estimate and predict the evolution of the carrier density and the resistivity, before the experiments are carried out. On the contrary, the theories with activation energy and with hopping rates do not give us that freedom, hence these other approaches cannot predict the evolution of the resistivity for different doping \textit{a priori}. In other words, these other approaches need to estimate the carrier density by other means to feed into their theory so as to predict the doping dependent resistivity. Apart from that, our resistivity theory captures the resistivity curve completely, from $T > 0$ to $T > T_C$ with a single equation, as presented by Eq.~(\ref{eq:7}). We do not have one equation for $T > T_C$, another for $T < T_C$ and another one for $T < T_{\rm{crossover}}$. Our theory captures all three mechanisms with a single equation, which will be very useful for experimental evaluations. However, this approach has not been developed to the extend where we can apply it to magnetic heterostructures to estimate the spin transmission probability as discussed by Egues~\cite{carlos}, Dai et al.~\cite{dai}, and Papp and Peeters~\cite{pee,pee2,pee3}.    

The proof that is given in the subsequent section is a form of proof-of-existence. This appendix is to prove that Eq.~(\ref{eq:100}) is mathematically and physically correct, regardless of the type of potential used. In order to prove this, we used the simplest case of 1D harmonic oscillator for convenience. Even if we were to use the hydrogenic wavefunction, we will still arrive at Eq.~(\ref{eq:100}) as derived in Ref.~\cite{andrew}. In mathematics, there are many techniques to prove the existence of a solution for a differential equation. Of course, one of the techniques is to \textit{solve} the equation. There are also the so-called existence and direct proofs. Now, to prove the validity of Eq.~(\ref{eq:100}) is quite easy because $\xi$ is also an eigenvalue and we did not touch the Hamilton operator. Hence, we are not required to solve Eq.~(\ref{eq:100}) in order to prove its validity. In other words, we can prove by means of constructive (existence) and/or direct proofs, by choosing a particular form of wavefunction with known solution and then calculate the total energy by comparison. In doing so, we will find that the total energy is always given by $E_0 \pm \xi$, as it should be. For example, for 1D harmonic oscillator, the known solution is a gaussian function, for 1D Dirac-delta potential, the known solution is an exponential function, for a 1D square well potential, the known solution is a sinusoidal function, for a 3D Coulomb potential in hydrogen atom, the known radial solution is an exponential function. Therefore, it depends on which system we are interested in and if we are interested in 1D harmonic oscillator, then we write the solution for Eq.~(\ref{eq:100}) in its gaussian form and then we derive the Schrodinger equation and its wavefunction in terms of $E_0 \pm \xi$. After that, we compare this new exponential wavefunction with the known solution for 1D harmonic oscillator. When we compare the constants, we will be able to show that the total energy is always given by $E_0 \pm \xi$.   

\subsection{1D harmonic oscillator: ionization energy as the eigenvalue}

The one-dimensional Hamiltonian of mass $m$ moving in the presence
of potential, $V(x)$ is given by (after making use of the linear
momentum operator, $\hat{p} = -i\hbar\partial^2/\partial x^2$)~\cite{griffiths5},

\begin {eqnarray}
\bigg[-\frac{\hbar^2}{2m}\frac{\partial^2}{\partial x^2} +
V(x)\bigg]\varphi &&= E\varphi, \nn \\&& = (E_{\rm{kin}} +
V_{\rm{pot}})\varphi, \label{eq:A1}
\end {eqnarray}

where $E$, $E_{\rm{kin}}$, $V(x)$ and $V_{\rm{pot}}$ denote the total energy,
kinetic energy, potential energy operator and the potential
energy, respectively.

We define,
\begin {eqnarray}
\pm\xi := E_{\rm{kin}} - E_0 + V_{\rm{pot}}, \label{eq:A2}
\end {eqnarray}

such that $\pm\xi$ is the energy needed for a particle to overcome
the bound state and the potential that surrounds it. The + sign for $\pm\xi$ is for the electron ($0 \rightarrow +\infty$) while the
$-$ sign is for the hole ($-\infty \rightarrow 0$). $E_{\rm{kin}}$ and
$E_0$ denote the total energy at $V_{\rm{pot}}$ = 0 and the energy at $T$
= 0, respectively, i.e., $E_{\rm{kin}}$ = kinetic energy. In physical
terms, $\xi$ is defined as the ionization energy. That is, $\xi$
is the energy needed to excite a particular electron to a finite distance, $r$, not necessarily $r \rightarrow \infty$.

On the other hand, using the above stated new definition
(Eq.~(\ref{eq:A2})) and the condition, $T = 0$ and $V_{\rm{pot}}$ = 0, we can rewrite the
total energy as

\begin {eqnarray}
E = E_{\rm{kin}} = E_0 \pm \xi.\label{eq:A5}
\end {eqnarray}

From Eq.~(\ref{eq:A2}) we have $E = E_0
\pm \xi$ = $E_{\rm{kin}} + V_{\rm{pot}}$, therefore

\begin {eqnarray}
\hat{H}\varphi = (E_0 \pm \xi)\varphi, \label{eq:A7}
\end {eqnarray}

where the total energy is
given by $E$ = $E_0 \pm \xi$ and $\varphi$ is the ionization energy based wavefunction while
$\hat{H}$ is the Hamilton operator.

\textbf{Proof}: Assume a solution for Eq.~(\ref{eq:A7}) at $n$ = 0
state (ground state) in the form of $\varphi_{n = 0}(x) = C
\exp[-ax^2]$ in order to be compared with the
standard harmonic oscillator wavefunction~\cite{griffiths5},

\begin {eqnarray}
&&\varphi_0(x) = \bigg[\frac{m \omega}{\pi \hbar
}\bigg]^{1/4}\exp\bigg[-\frac{m \omega}{2 \hbar}x^2\bigg],
\label{eq:A8}
\end {eqnarray}

where $\hbar$ $=$ $h/2\pi$, $h$ is Planck's constant and $\omega$ is the frequency. Therefore, we obtain $\ln \varphi(x) = \ln C - ax^2 \ln e$,
$\frac{1}{\varphi(x)} \frac{\partial \varphi(x)}{\partial x} =
-2ax$ and $\frac{\partial^2\varphi(x)}{\partial x^2} =
2a\varphi(x)[2ax^2-1]$. On the other hand, we can rewrite
Eq.~(\ref{eq:A7}) to obtain

\begin {eqnarray}
&&\frac{\partial^2\varphi}{\partial x^2} = -\frac{2m}{\hbar^2}[E_0
\pm \xi]\varphi, \label{eq:A12}
\end {eqnarray}

where $E$ and $E_0$ for a given system range from $+\infty$ to 0 for
electrons and 0 to $-\infty$ for holes which explains
the $\pm$ sign in $\xi$. Using Eqs.~(\ref{eq:A1}),~(\ref{eq:A7})
and,~(\ref{eq:A12}), we obtain $a = \frac{m}{\hbar^2}(E_0 \pm
\xi)$. Normalizing $\varphi$ gives

\begin {eqnarray}
&& 1 = \int_{-\infty}^{+\infty}|\varphi(x)|^2 dx = 2|C|^2
\int_0^{+\infty}e^{-2ax^2} dx \nn\\&& = 2|C|^2 \frac{1}{2}
\left[\frac{\pi}{2a}\right]^{1/2} \label{eq:A15}
\end {eqnarray}

and $C = \left[\frac{2m(E_0 \pm \xi)}{\pi \hbar^2 }\right]^{1/4}$.
Consequently,

\begin {eqnarray}
\varphi(x) = \left[\frac{2m(E_0 \pm \xi)}{\pi \hbar^2
}\right]^{1/4}\exp\bigg[-\frac{m}{\hbar^2}(E_0 \pm
\xi)x^2\bigg].\label{eq:A17}
\end {eqnarray}

Now, we compare Eq.~(\ref{eq:A17}) with Eq.~(\ref{eq:A8}). In
doing so, we can show that the ground state energy is,

\begin {eqnarray}
\frac{1}{2}\hbar \omega = E_0 \pm \xi, \label{eq:A18}
\end {eqnarray}

either from equating $\big[m \omega/\pi \hbar\big]^{1/4} = \big[2m(E_0 \pm
\xi))/\pi \hbar^2\big]^{1/4}$ or  $-m
x^2\omega/2 \hbar$ = $-m(E_0 \pm \xi)x^2/\hbar^2$.

\subsection{Expectation value for V(x)}

In this section, we will show that the total energy, $E_0 \pm \xi$ is a function of the potential energy. That is, using Eqs.~(\ref{eq:A1}),~(\ref{eq:A2})
and~(\ref{eq:A7}), we will show the potential energy can be written in terms of
$E_0 \pm \xi$. For example, from Eq.~(\ref{eq:A17}) the harmonic
oscillator Schrodinger Eq. can be shown as

\begin {eqnarray}
-\frac{\hbar^2}{2m}\frac{\partial^2\varphi}{\partial x^2} =
\bigg[\frac{2m}{\hbar^2}(E_0 \pm \xi)^2x^2 - (E_0 \pm
\xi)\bigg]\varphi. \label{eq:A19}
\end {eqnarray}

Therefore, the potential energy is given by

\begin {eqnarray}
V(x) = \frac{2m}{\hbar^2}(E_0 \pm \xi)^2x^2. \label{eq:A20}
\end {eqnarray}

\textbf{Proof}: From Eq.~(\ref{eq:A20}), we can write

\begin {eqnarray}
\hat{H} = \frac{1}{2m}\bigg[\hat{p}^2 + \bigg(\frac{2m}{\hbar}(E_0
\pm \xi)x\bigg)^2\bigg]. \label{eq:A21}
\end {eqnarray}

Therefore, the ladder operator can be written as

\begin {eqnarray}
a_{\pm} = A\bigg[\mp i\hat{p} + \bigg(\frac{2m}{\hbar}(E_0 \pm
\xi)\bigg)x\bigg]. \label{eq:A22}
\end {eqnarray}

$A$ is a factor that will be used to derive the expectation value of $V(x)$. Taking $K = \frac{2m}{\hbar}(E_0 \pm \xi)$, we obtain

\begin {eqnarray}
a_{\pm} = A[\mp i\hat{p} + Kx]. \label{eq:A23}
\end {eqnarray}

Since the commutation relation~\cite{griffiths5}, $[x,\hat{p}] =
i\hbar$, then

\begin {eqnarray}
[a_-,a_+] = 2A^2K\hbar = -[a_+,a_-].\label{eq:A24}
\end {eqnarray}

Using Eq.~(\ref{eq:A21}), we get

\begin {eqnarray}
&&a_-a_+ = A^2(2m\hat{H} + K\hbar), \label{eq:A25}\\&& \hat{H} =
\frac{1}{2m}\left[\frac{a_{\mp}a_{\pm}}{A^2} \mp K\hbar\right].
\label{eq:A26}
\end {eqnarray}

Consequently, we can show that

\begin {eqnarray}
\hat{H}(a_+\varphi) &=& \frac{1}{2m}\left[\frac{a_+a_-}{A^2} +
K\hbar\right](a_+\varphi) \nn \\&=& (E_0 \pm \xi)\varphi. \label{eq:A28}
\end {eqnarray}

Subsequently, $\hat{H}(a_-\varphi) = \left[E_0 \pm \xi - 2(E_0 \pm
\xi)\right]a_-\varphi$. Applying the condition~\cite{griffiths5}
for the ground state, $\varphi_{n=0}$ such that

\begin {eqnarray}
a_-\varphi_{n=0} = 0, \label{eq:A30}
\end {eqnarray}

will lead us to

\begin {eqnarray}
E_{n=0}\varphi_{n=0} = \hat{H}\varphi_{n=0} = (E_0 \pm \xi)\varphi_{n=0} .\label{eq:A31}
\end {eqnarray}

Recall that $E_0$ is the energy at $T$ = 0 and $E_{n=0}$ denotes
the energy for the $n = 0$ state. Finally, utilizing
Eqs.~(\ref{eq:A31}) and~(\ref{eq:A28}), we obtain

\begin {eqnarray}
E_n &=& (1 + 2n)(E_0 \pm \xi).\label{eq:A32}
\end {eqnarray}

Subsequently, we find that (using
Eqs.~(\ref{eq:A7}),~(\ref{eq:A26}) and~(\ref{eq:A32}))

\begin {eqnarray}
&&a_+a_-\varphi_n = n2A^2K\hbar \varphi_n, \nn\\&& a_-a_+\varphi_n
= (n+2A^2K\hbar)\varphi_n \Leftrightarrow [a_-,a_+] = 2A^2K\hbar.
\nn
\end {eqnarray}

Now, using the identity~\cite{griffiths5}

\begin {eqnarray}
\int_{-\infty}^{\infty}(f^*)(a_{\pm}g)dx =
\int_{-\infty}^{\infty}(a_{\mp}f^*)(g) dx, \nn
\end {eqnarray}

We find

\begin {eqnarray}
\int_{-\infty}^{\infty}(a_+\varphi_n)^*(a_+\varphi_n)dx &&= \nn 
\int_{-\infty}^{\infty}(a_-a_+\varphi_n)^*\varphi_n dx \\&& \nn =
(n+2A^2K\hbar)\int_{-\infty}^{\infty}|\varphi_n|^2 dx, \nn
\end {eqnarray}

\begin {eqnarray}
\int_{-\infty}^{\infty}(a_-\varphi_n)^*(a_-\varphi_n)dx &&= \nn
\int_{-\infty}^{\infty}(a_+a_-\varphi_n)^*\varphi_n dx \\&& \nn =
n2A^2K\hbar \int_{-\infty}^{\infty}|\varphi_n|^2 dx. \nn
\end {eqnarray}

On the other hand, we have 

\begin {eqnarray}
a_{\pm} \varphi_n \propto \varphi_{n \pm 1} \Rightarrow a_{\pm} \varphi_n = \gamma^{\pm}_n \varphi_{n \pm 1},\nn 
\end {eqnarray}

where $\gamma^{+}_n$ and $\gamma^{-}_n$ are the proportionality factors, which can be determined from 

\begin {eqnarray}
\int_{-\infty}^{\infty}(a_{\pm}\varphi_n)^*(a_{\pm}\varphi_n)dx = \nn 
|\gamma^{\pm}_n|^2\int_{-\infty}^{\infty}|\varphi_{n \pm 1}|^2 dx, \nn
\end {eqnarray}

hence, we can now write $a_+\varphi_n =
(n+2A^2K\hbar)^{1/2}\varphi_{n+1}$ and

\begin {eqnarray}
a_-\varphi_n = (n2A^2K\hbar)^{1/2}\varphi_{n-1}.\label{eq:A39}
\end {eqnarray}

We can rearrange Eq.~(\ref{eq:A23}) to get

\begin {eqnarray}
x^2 = \frac{(a_++a_-)^2}{4A^2K^2}.\label{eq:A40}
\end {eqnarray}

As a consequence, (from Eqs.~(\ref{eq:A23}) and~(\ref{eq:A40}), also invoking the orthogonality in the last step)

\begin {eqnarray}
\langle V(x)\rangle &=& \bigg\langle \frac{2m}{\hbar^2}(E_0 \pm
\xi)^2x^2\bigg\rangle \nn \\&=& \frac{2A^2K\hbar(1+n) +
n}{8A^2m}.\nn \\&& \label{eq:A41}
\end {eqnarray}

From Eq.~(\ref{eq:A32}), we know that

\begin {eqnarray}
\langle V(x)\rangle = \frac{1}{2}(1 + 2n)(E_0 \pm
\xi),\label{eq:A42}
\end {eqnarray}

the other half is due to kinetic energy~\cite{griffiths5}. Putting
Eqs.~(\ref{eq:A41}) and~(\ref{eq:A42}) together leaves us with

\begin {eqnarray}
\frac{2A^2K\hbar(1+n) + n}{8A^2m} = \frac{1}{2}(1 + 2n)(E_0 \pm
\xi),\label{eq:A43}
\end {eqnarray}

Therefore $A^2 = \frac{1}{2K\hbar}$. Hence, the commutation
relation given in Eq.~(\ref{eq:A24}) can be rewritten as

\begin {eqnarray}
[a_-,a_+] = 2A^2K\hbar = 1 = -[a_+,a_-].\label{eq:A47}
\end {eqnarray}

As a result of this, indeed the potential energy is given in terms
of $E_0 \pm \xi$ from Eq.~(\ref{eq:A42}).


\begin{references}

\bibitem{igor} I. Zutic, J. Fabian and S. D. Sarma, Rev. Mod. Phys. 76 (2004) 323.

\bibitem{munekata1} H. Munekata, H. Ohno, S. von Molnar, A. Segmuller, L. L. Chang and L. Esaki, Phys. Rev. Lett. 63 (1989) 1849.

\bibitem{gub} R. von Helmolt, J. Wecker, B. Holzapfel, L. Schultz and K. Samwer, Phys. Rev. Lett. 71 (1993) 2331.

\bibitem{mus} M. Eginligil, G. Kim, Y. Yoon, J. P. Bird, H. Luo and B. D. McCombe, Physica E 40 (2008) 2104.

\bibitem{kana} I. Kanazawa, Physica E 40 (2007) 277.

\bibitem{sami} L. Saminadayar, P. Mohanty, R. A. Webb, P. Degiovanni and C. Bauerle, Physica E 40 (2007) 12.

\bibitem{jaya} T. Jayasekera, N. Goel, M. A. Morrison and K. Mullen, Physica E 34 (2006) 584.

\bibitem{esch13} A. van Esch, L. van Bockstal, J. de Boeck, G. Verbanck, A. S. van Steenbergen, P. J. Wellmann, B. Grietens, R. Bogaerts, F. Herlach and G. Borghs, Phys. Rev. B 56 (1997) 13103.

\bibitem{t-omiya} T. Omiya, F. Matsukura, T. Dietl, Y. Ohno, T. Sakon, M. Motokawa and H. Ohno, Physica E 7 (2000) 976.

\bibitem{lutt} J. M. Luttinger and W. Kohn, Phys. Rev. 97 (1955) 869.

\bibitem{jung2} T. Jungwirth, J. Sinova, J. Macek, J. Kucera, and A. H. MacDonald, Rev. Mod. Phys. 78 (2006) 809.

\bibitem{hwang} E. H. Hwang E H and S. D. Sarma, Phys. Rev. B 72 (2005) 35210.

\bibitem{lopez} M. P. Lopez-sancho and L. Brey, Phys. Rev. B 68 (2003) 113201.

\bibitem{sen} C. Sen, G. Alvarez, H. Aliaga and E. Dagotto, Phys. Rev. B 73 (2006) 224441.

\bibitem{mayr} M. Mayr, A. Moreo, J. A. Verges, J. Arispe, A. Feiquin and E. Dagotto, Phys. Rev. Lett. 86 (2000) 135.

\bibitem{dietl} T. Dietl, Physica E 35 (2006) 293.

\bibitem{arulsamy2} A. D. Arulsamy, Physica C 356 (2001) 62, arXiv:cond-mat/0402153

\bibitem{arulsamy3} A. D. Arulsamy, Phys. Lett. A 300 (2002) 691.

\bibitem{arulsamy7} A. D. Arulsamy in \textit{Superconductivity
research at the leading edge} (ed Lewis, P. S.) 45 (Nova Science Publishers, New York, 2004).

\bibitem{arulsamy8} A. D. Arulsamy, Phys. Lett. A 334 (2005) 413.

\bibitem{anderson} P. W. Anderson, Science 177 (1972) 393.

\bibitem{andrew} A. D. Arulsamy, arXiv:physics/0702232v9; arXiv:0807.0745.

\bibitem{burch} K. S. Burch, D. B. Shrekenhamer, E. J. Singley, J. Stephens, B. L. Sheu, R. K. Kawakami, P. Schiffer, N. Samarth, D. D. Awschalom and D. N. Basov, Phys. Rev. Lett. 97 (2006) 87208.

\bibitem{cou} O. D. D. Couto, J. Rudolph, F. Likawa, R. Hey and P. V. Santos, Physica E 40 (2008) 1797.

\bibitem{kaes} B. Kaestner, J. Wunderlich, T. Jungwirth, J. Sinova, K. Nomura and A. H. MacDonald, Physica E 34 (2006) 47.

\bibitem{beth} H. A. Bethe and E. E. Salpeter, \textit{Quantum mechanics of one- and two-electron atoms} (Springer-Verlag, Berlin, 1957).

\bibitem{murakami} S. Murakami, N. Nagaosa and S. C. Zhang, Science 301 (2003) 1348. 

\bibitem{tinbergen3} Tineke Van Peski-Tinbergen and A. J. Dekker, Physica 29 (1963) 917.

\bibitem{kasuya4} T. Kasuya, Prog. Theor. Phys. 16 (1956) 58.

\bibitem{matsu} F. Matsukura, H. Ohno, A. Shen and Y. Sugawara, Phys. Rev. B 57 (1998) R2037.

\bibitem{web28} M. J. Winter $<$http://www.webelements.com$>$.

\bibitem{oiwa} A. Oiwa, S. Katsumoto, A. Endo, M. Hirasawa, Y. Iye, 
H. Ohno, F. Matsukura, A. Shen and Y. Sugawara, Solid State Commun. 103 (1997) 209.

\bibitem{priya} P. Mahadevan and A. Zunger, Phys. Rev. B 68 (2003) 75202. 

\bibitem{park3} Y. D. Park, A. T. Hanbicki, S. C. Erwin, C. S. Hellberg, J. M. Sullivan, J. E. Matson, T. F. Ambrose, A. Wilson, G. Spanos and B. T. Jonker, Science 295 (2002) 651.

\bibitem{mahendiran2} R. Mahendiran, S. K. Tiwary, A. K. Raychaudhuri, T. V. Ramakrishnan, R. Mahesh, N. Rangavittal and C. N. R. Rao, Phys. Rev. B 53 (1996) 3348.

\bibitem{carlos} J. C. Egues, Phys. Rev. Lett. 80 (1998) 4578.

\bibitem{dai} N. Dai, H. Luo, F. C. Zhang, N. Samarth, M. Dobrowolska and J. K. Furdyna, Phys. Rev. Lett. 67 (1991) 3824.

\bibitem{pee} G. Papp and F. M. Peeters, Phys. Stat. Sol. (b) 241 (2004) 222.

\bibitem{pee2} G. Papp and F. M. Peeters, Appl. Phys. Lett. 78 (2001) 2184.

\bibitem{pee3} G. Papp and F. M. Peeters, Appl. Phys. Lett. 79 (2001) 3198.

\bibitem{griffiths5} D. J. Griffiths \textit{Introduction to quantum mechanics} (Prentice-Hall, New Jersey, 1995).

\end{references}
\end{document}